\documentclass[aps,prev,twocolumn,preprintnumbers,floatfix,nofootinbib]{revtex4-1}
\pdfoutput=1
\usepackage{graphicx}
\usepackage{bm}
\usepackage{times}
\usepackage{hyperref}
\usepackage{slashed}
\usepackage{color}
\usepackage{appendix}
\usepackage{mathtools}
\usepackage{epsfig}
\usepackage{dcolumn}
\usepackage{textcomp}
\usepackage{subcaption}

\begin{document} 

\title{ Ultraheavy Dark Matter and WIMPs Production aided by Primordial Black Holes}

\author{Giorgio Arcadi$^{1,2}$}
\email{giorgio.arcadi@unime.it}
\author{Manfred Lindner$^{3}$}
\email{lindner@mpi-hd.mpg.de}
\author{Jacinto P. Neto$^{1,3,4,5}$}
\email{jacinto.neto.100@ufrn.edu.br}
\author{Farinaldo S. Queiroz$^{4,5,6}$}
\email{farinaldo.queiroz@ufrn.br}

\affiliation{$^1$Dipartimento di Scienze Matematiche e Informatiche, Scienze Fisiche e Scienze della Terra, Universita degli Studi di Messina, Viale Ferdinando Stagno d’Alcontres 31, I-98166 Messina, Italy}
\affiliation{$^2$INFN Sezione di Catania, Via Santa Sofia 64, I-95123 Catania, Italy}
\affiliation{$^3$ Max Planck Institut für Kernphysik, Saupfercheckweg 1, D-69117 Heidelberg, Germany}
\affiliation{$^4$ Departamento de F\'isica, Universidade Federal do Rio Grande do Norte, 59078-970, Natal, RN, Brasil}
\affiliation{$^5$ International Institute of Physics, Universidade Federal do Rio Grande do Norte, Campus Universitário, Lagoa Nova, Natal-RN 59078-970, Brazil}
\affiliation{$^6$ Millennium Institute for Subatomic Physics at High-Energy Frontier (SAPHIR),
Fernandez Concha 700, 7550196, Santiago, Chile}

\begin{abstract}
\noindent

The unitary bound restricts thermal relics to be lighter than $100$~TeV. This work investigates the production of ultraheavy dark matter and WIMPs in the presence of primordial black holes. Firstly, we describe how Hawking evaporation can produce ultraheavy dark matter with masses above $10^{12}$~GeV in radiation and matter-domination eras. Later, we assess how primordial black holes that induce a non-standard cosmology impact the predicted relic density of a thermal relic and explore the interplay between them, considering the restrictions arising from entropy injection due to the evaporation of primordial black holes. Considering a concrete B-L model, where the dark matter is a Dirac particle, we obtain the correct relic density for various freeze-out scenarios and show that a dark matter particle can nicely reproduce the correct relic density in agreement with current limits with masses above the 10 TeV scale. Hence, this work strengthens the continuous search for heavy dark matter particles. 
\end{abstract}


\maketitle
\flushbottom

\section{\label{Intro} Introduction}
We have collective evidence for dark matter rising from galaxy formation, Cosmic Microwave Background (CMB), gravitation lensing, among others \cite{Bertone:2004pz}. CMB data requires a cold dark matter component, with relic density $\Omega_{\textrm{DM}} h^2 =  0.1200 \pm 0.0012$ within 68\% C.L. \cite{Planck:2018vyg}. The absence of non-gravitational signals keeps the nature of dark matter thus far unknown \cite{LZ:2022ufs,XENON:2023sxq}. Dark matter is often interpreted as elementary particles or primordial black holes. Different production mechanisms were proposed to explain the dark matter genesis. In the context of elementary particles, the most appealing non-gravitational dark matter production is the thermal freeze-out  (see e.g. \cite{Arcadi:2024ukq} for a recent review). In the thermal freeze-out mechanism, the dark matter particles is assumed to be in thermal equilibrium with Standard Model (SM) particles along the thermal history of the universe, until the chemical decoupling. As the interactions that control freeze-out are similar to the ones that govern direct and indirect detection signals, thermal relics are subject to restrictive bounds from direct detection and indirect detection experiments \cite{LZ:2023lvz,XENON:2023cxc,CTA:2020qlo}. In light of that, alternative dark matter production mechanisms have been investigated \cite{Matos:2023usa,Hochberg:2014dra}.

In an orthogonal direction, Primordial Black Holes (PBHs) have been revived as viable dark matter candidates in light of Gravitational Waves (GW) related discoveries and revisited bounds \cite{Villanueva-Domingo:2021spv,Oncins:2022ydg}. An interesting interplay between PBHs and dark matter particles has attracted attention in the form of Hawking evaporation \cite{Hawking:1974rv, Hawking:1975vcx, Ema:2018ucl, Mambrini:2021zpp, Bernal:2022oha}. PBHs can yield the observed dark matter relic density alone or act as an additional production mechanism \cite{Kitabayashi:2022fqq, Gondolo:2020uqv, Chaudhuri:2023aiv}.

In this work, we further investigate the impact of ultralight PBHs on dark matter production \cite{Carr:1974nx, Bernal:2021akf, Cirelli:2024ssz, Profumo:2024fxq}. Although we consider PBHs that evaporate before Big Bang Nucleosynthesis (BBN), they produce dark matter particles that survive until today, the so-called Hawking relics \cite{Shallue:2024hqe}. Moreover, PBHs may trigger an early matter-dominated period depending on their initial abundance, which above a critical value becomes unavoidable \cite{Gondolo:2020uqv}. 

PBHs have been extensively studied in the literature since they can help us understand still current mysteries, such as dark matter and matter-antimatter asymmetry. They can give rise to the observed dark matter density alone \cite{Kitabayashi:2022fqq, Gondolo:2020uqv, Chaudhuri:2023aiv} or together with other mechanisms  \cite{Bernal:2020bjf, Bernal:2021yyb, Bernal:2021bbv, Mazde:2022sdx, Cui:2024uwk}. Furthermore, PBHs can be the source of non-standard cosmological histories, such as an early matter-dominated universe \cite{Arias:2019uol,Arcadi:2021doo,Arcadi:2024jzv,Arias:2023wyg}.

Motivated by this exciting journey involving dark matter and PBHs, in this work, we assess a mixed scenario of two-component dark matter, in which an ultraheavy dark matter is gravitationally produced via Hawking evaporation and a WIMP candidate that rises from non-standard freeze-out production led by PBHs \cite{Boucenna:2017ghj, Adamek:2019gns, Bertone:2019vsk, Carr:2020mqm}. We exploit the effects of a PBH domination on the freeze-out production of a vector-like Dirac fermion WIMP embedded in a minimal $U(1)_{B-L}$ extension. We demonstrate that the entropy injection is small and provide approximate analytical solutions for the Boltzmann equations that govern the dark matter abundance.
 
The paper is organized as follows: We review the fate of these ultralight PBHs and their generated particle spectrum in Sec.~\ref{sec:PBHs}. In Sec.~\ref{sec:DMproduc}, we derive the relevant equations concerning dark matter genesis via Hawking evaporation and standard freeze-out. Also, we obtain the approximate analytical solutions for the non-standard freeze-out scenarios and discuss the constraints on the PBH parameter space. In Sec.~\ref{sec:models}, we present the WIMP dark matter model; In Sec.~\ref{sec:Results}, we derive the relevant bounds. Finally, in Sec.~\ref{sec:Conclusions} we draw our conclusions.   

\section{\label{sec:PBHs} Primordial Black Holes' death}
PBHs could have been formed in the early Universe through various mechanisms. In this work, we will not dive into the formation process and invoke the presence of a mechanism giving rise to a monochromatic mass distribution. For more details see \cite{Carr:1975qj, Masina:2020xhk, Cheek:2021odj, Kleban:2023ugf}. We assume Schwarzschild's black holes, whose properties are described solely by their mass, $M_{\rm BH}$. 
These PBHs arose with an initial mass proportional to the particle horizon mass during a radiation-dominated era with plasma temperature $T=T_{\rm{in}}$ \cite{Carr:2020gox},
\begin{equation}
    M_{\rm{in}} \equiv M_{\rm{BH}} (T_{\rm{in}})= \frac{4 \pi \gamma}{3}\frac{\rho_{\rm rad}(T_{\rm in})}{H^{3}(T_{\rm in})}, \label{eq:Min}
\end{equation}
where $\gamma = \omega^{3/2} \approx 0.2$ is a dimensionless parameter related to the gravitational collapse\footnote{$\omega \equiv p/\rho = 1/3$ stands for the parameter of the equation of state in a radiation-dominated epoch.}, and the radiation energy density and the Hubble parameter are, respectively,
\begin{align}
    \rho_{\rm rad}(T) &= \frac{\pi^2}{30}g_\star(T)T^4, \quad {\rm and} \label{eq:raddensity} \\
    H(T) &= \sqrt{ \frac{\rho(T)}{3 M^2_P}} = \frac{\pi}{3}\sqrt{\frac{g_\star(T)}{10}}\frac{T^2}{M_{\rm P}} , \label{eq:Hubble}
\end{align}
with the Planck mass $M_{\rm P} = \sqrt{1/ (8\pi G)} = 2.435 \times 10^{18}$~GeV $\simeq 4.334 \times 10^{-6}$~g, and $g_\star (T)$ is the total number of relativistic degrees of freedom at the primordial plasma temperature $T$. 

Hawking radiation processes offer a connection between gravitational and quantum phenomena \cite{Hawking:1974rv, Hawking:1975vcx}. Such an effect forces a black hole to lose its mass through the emission of particle species lighter than its event horizon temperature, $T_{\rm BH}$, which in turn yields a thermal radiation spectrum \cite{Page:1976df, Baldes:2020nuv,Cheek:2021odj} with temperature related to its mass as follows \cite{Hawking:1975vcx},
\begin{equation}
    T_{\rm BH} = \frac{M_{\rm P}^2}{M_{\rm BH}}.
\end{equation}
The energy spectrum of the $i$-th emitted particle by a Schwarzschild black hole is \cite{Page:1976df, Gondolo:2020uqv, Bernal:2021yyb},
\begin{equation}\label{eq:totenergyperunitarea}
    \frac{d^2 \mathcal{U}_j (E,t)}{dt \, dE} = \frac{g_j}{8 \pi^2} \frac{E^3}{\exp(E/T_{\rm BH}) \pm 1}, \begin{cases}
        +, \,\,{\rm Fermions} \\
        -, \,\,{\rm Bosons}
    \end{cases}
\end{equation}
where $\mathcal{U}_j (E,t)$ stands for the total released energy per unit area, $E$ accounts for the energy of the emitted particle, and $g_j$ is the number of degrees of freedom of the species $j$. Hawking radiation makes a black hole lose mass at a rate \cite{Bernal:2021yyb}, 
\begin{align}
    \frac{dM_{\rm BH}}{dt} &= -4 \pi r_S^2 \sum_{j} \int_{0}^{\infty} \frac{d^2 \mathcal{U}_j (E,t)}{dt\,dE} dE \nonumber \\
    &= - \frac{\pi g_\star(T_{\rm BH})}{480} \frac{M_{\rm P}^4}{M_{\rm BH}^2} \label{eq:PBHMassevolution},
\end{align}
where $r_S = M_{\rm BH}/(4\pi M_{\rm P}^2)$ is the Schwarzschild radius. Finally, integrating the equation above from an initial mass $M_{\rm in}$ at initial time $t_{\rm in}$ up to a mass $M_{\rm BH}(t)$ at time $t$, we find the time evolution of the black hole mass,
\begin{equation}
    M_{\rm BH}(t) = M_{\rm in} \left( 1 - \frac{t - t_{\rm in}}{\tau}\right)^{1/3},
\end{equation}with the black hole {\it lifetime} being,
\begin{equation}
    \tau = \frac{160}{ \pi g_\star(T_{\rm in})}\frac{M_{\rm in}^3}{M_{\rm P}^4}.\label{eq:lifetimepbh}
\end{equation}
Using Eq.~\eqref{eq:Min}, and equating Eq.~\eqref{eq:Hubble} to $H = 1/(2 \tau)$, assuming $g_\star = 106.8$ and $\gamma = 0.2$, we can show that $t_i/\tau \simeq 7.8\times 10^{-12}({\rm g}/M_{\rm in})^{2}$, hence, $\tau \gg t_{\rm in}$. If PBH never dominates the total energy density of the Universe before they completely evaporate, we can follow a similar approach to find that the temperature of the thermal bath right after complete evaporation for a radiation-dominated universe $T^{\rm rad}_{\rm ev} \simeq T(\tau)$ is,
\begin{equation}
    T_{\rm ev}^{\rm rad} \simeq \frac{\sqrt{3}}{4}\left( \frac{g_{\star}(T_{\rm in})}{45}\right)^{1/4}\left(\frac{M_{\rm P}^{5}}{M_{\rm in}^3}\right)^{1/2}.
\end{equation}

However, if PBHs come to dominate, we equal Eq.~\eqref{eq:Hubble} to $H = 2/(3 \tau)$ (Friedmann solution to matter-domination), and obtain a different evaporation temperature,
\begin{equation}\label{eq:Tevbar}
    T_{\rm ev}^{\rm mat} \simeq \left( \frac{g_{\star}(T_{\rm in})}{640}\right)^{1/4} \left(\frac{M_{\rm P}^5}{M_{\rm in}^3}\right)^{1/2} \equiv T_{\rm ev}.
\end{equation}

The Hawking spectrum for a given particle species $j$ can be written in terms of its energy spectrum in Eq.~\eqref{eq:totenergyperunitarea} \cite{Gondolo:2020uqv, Bernal:2020bjf},
\begin{align}
    \frac{d^2 \mathcal{N}_j}{dt\,dE} &= \frac{4\pi r_S^2}{E}\frac{d^2 \mathcal{U}_j}{dt\,dE} \nonumber \\
    & = \frac{g_j\,r_S^2}{2\pi}\frac{E^2}{\exp(E/T_{\rm BH}) \pm 1}.\label{eq:particlespectrumpbh}
\end{align}

Integrating over the lifetime $\tau$ and the total energy $E$, we obtain \cite{Baldes:2020nuv, Gondolo:2020uqv},
\begin{align}
    \mathcal{N}_j &= \mathcal{C}_n \frac{120 \zeta(3)}{\pi^3}  \frac{g_j}{g_\star(T_{\rm in})} \left( \frac{M_{\rm in}}{M_{\rm P}}\right)^2, \quad m_j < T_{\rm in} \label{eq:Ntot1}\\
    \mathcal{N}_j &= \mathcal{C}_n \frac{15 \zeta(3)}{8 \pi^5}  \frac{g_j}{g_\star(T_{\rm in})} \left( \frac{M_{\rm P}}{m_j}\right)^2, \quad m_j > T_{\rm in}, \label{eq:Ntot2}
\end{align}
the total number of particle species $j$ emitted by the primordial black holes during its evaporation, where $\mathcal{C}_n = 1$ or $3/4$ for bosonic or fermionic species, respectively.

Our goal was to obtain the production of species due to Hawking evaporation, which is the foundation for dark matter production via Primordial Black Holes as we discuss in the next section.

\section{\label{sec:DMproduc} Dark matter genesis}

In this section, we will investigate the dark matter production from two sources, namely Hawking evaporation and thermal freeze-out. As a gravitational process, Hawking evaporation can produce ultraheavy dark matter particles, beyond the unitarity limit $m_{\rm DM} \gg 10^5$~GeV. On the other hand, the freeze-out mechanism can only produce dark matter species below  $10^5 \, {\rm GeV}$. In this work, we will explore the production of ultraheavy dark matter particles from PBHs and the WIMP production in a non-standard freeze-out scenario caused by PBHs. In other words, the overall abundance of dark matter is given by,
\begin{align}
    \Omega_{\textrm{DM}}h^2 = (\Omega_{\rm DM}^{\textrm{FO}} +\Omega_{\text{DM}}^{\textrm{BH}})h^2 \simeq 0.12.
\end{align}

As the dark matter abundance has two sources. It is useful to connect them via the dark matter Yield, $Y_{DM}=n_{DM}/s$, which is the dark matter comoving number density, with $n_{\rm DM}$ being the dark matter number density. The Yield is related to the  dark matter abundance through, 
\begin{equation}
    \Omega_{\rm DM} = \frac{\rho_{{\rm DM}, 0}}{\rho_{\rm c}} =  \frac{s_0}{\rho_{\rm c}} m_{\rm DM} Y_{\rm DM},
\end{equation}
where the subscript $0$ means its value today. Plugging the numbers for the entropy density (s) and critical density,
\begin{align}
    s_0 &= \frac{2891.2}{{\rm cm}^3}, \,\,\,{\rm and} \\
    \rho_{\rm c} &=  1.0537 \times 10^{-5}\,h^2\, \frac{{\rm GeV}}{{\rm cm}^3},
\end{align} we obtain,

\begin{equation}\label{eq:relicdensity}
    \Omega_{\rm DM} h^2 = 2.74385 \times 10^8 \,\frac{m_{\rm DM}}{\rm GeV} \,Y_{\rm DM}.
\end{equation}

With this information, we can explore the interplay between Hawking's evaporation and the freeze-out mechanism in the next section.

\subsection{\label{PBHev} Hawking evaporation}
We dedicate this section to introducing the genesis of dark matter through Hawking evaporation of PBHs. We discuss the production of Hawking relics when the universe is always dominated by SM radiation and when PBH domination leads to an early matter-dominated period. We avoid the production of Planck relics assuming that the Hawking evaporation does not end at $T_{\rm BH} \sim M_{\rm P}$.

\paragraph{Radiation-dominated era} We obtain the dark matter yield today $Y_{\textrm{DM}} \equiv n_{\textrm{DM}}(T_0)/s(T_0)$ tracing the comoving number density using entropy conservation back to the time of PBH formation \cite{Gondolo:2020uqv},
\begin{equation}\label{eq:PBHYield1}
    Y_{\textrm{DM}} = \frac{n_\textrm{DM}(T_{{\rm ev}}^{\rm rad})}{s(T_{{\rm ev}}^{\rm rad})} = N_{{\rm DM}} \frac{n_{\rm BH}(T_{{\rm in}})}{s(T_{{\rm in}})}, 
\end{equation}
where $n_{\textrm{DM, BH}}(T)$ is the number density of dark matter particles or primordial black holes. Moreover, $N_\textrm{DM}$ is the total number of dark matter particles emitted via Hawking evaporation given by Eqs.~\eqref{eq:Ntot1} and \eqref{eq:Ntot2}, 
and $s(T)$ is the entropy density,
\begin{equation}\label{eq:entropydensity}
    s(T) = \frac{2\pi^2}{45} g_{\star,s}(T)\,T^3,
\end{equation}
where $g_{\star, s}$ is the number of relativistic degrees of freedom contributing to the SM entropy. For PBHs forming during a radiation-dominated epoch, we can write the energy density as follows,
\begin{equation}\label{eq:rhobh}
    \rho_{{\rm BH}}(T) = \beta\, \rho_{{\rm rad}}(T), \quad {\rm with} \quad \beta \equiv \frac{\rho_{{\rm BH}}(T_{{\rm in}})}{\rho_{{\rm rad}}(T_{{\rm in})}}.
\end{equation}
Hence, $\beta$ accounts for the fraction of PBHs compared to the SM radiation energy density at the time of formation or temperature $T_{\rm in}$. We can use $\rho_{\rm BH}(T) = M_{\rm BH}(T) n_{\rm BH}(T)$ to write,
\begin{equation}
    \beta = M_{\rm BH}(T_{\rm in}) \frac{n_{\rm BH}(T_{\rm in})}{\rho_{\rm rad}(T_{\rm in})} = M_{\rm in} \frac{n_{\rm BH}(T_{\rm in})}{\rho_{\rm rad}(T_{\rm in})}.
\end{equation}
We anticipate that $\beta$ is a free parameter in this work. Rewriting it in terms of $n_{\rm BH}(T_{\rm in})$  and plugging it into Eq.~\eqref{eq:PBHYield1} to get,
\begin{equation}
    Y^{\rm RD}_{\rm DM} = \beta \,\frac{N_{\rm DM}}{M_{\rm BH}} \frac{\rho_{\rm rad}(T_{\rm in})}{s(T_{\rm in})} = \frac{3}{4}\frac{g_\star (T_{\rm in})}{g_{\star, s}(T_{\rm in})} \beta\,N_{\rm DM}\frac{T_{\rm in}}{M_{\rm in}},
\end{equation}
where we used Eq.~\eqref{eq:raddensity} and \eqref{eq:entropydensity}. From Eq.~\eqref{eq:Min} we find,
\begin{equation}\label{eq:Tin}
     T_{\rm in} = \sqrt{3\gamma}\left(\frac{160}{g_{\star}(T_{\rm in})}\right)^{1/4} \left(\frac{M_{\rm P}^3}{M_{\rm in}}\right)^{1/2}.
\end{equation}

Since we are interested in ultralight PBHs, $M_{\rm in} \in \{1 - 10^{8}\}$~g, 
there is no much loss of generality to assume $g_\star (T_{\rm in})= g_{\star, s} (T_{\rm in})$. Substituting Eq.~\eqref{eq:Tin} into $Y_{\rm DM}$, we obtain \cite{Gondolo:2020uqv},
\begin{equation}\label{eq:yieldrad}
    Y^{\rm RD}_{\rm DM} = \sqrt{3 \gamma}\left(\frac{405}{8\,g_\star(T_{\rm in})}\right)^{1/4}\beta\,N_{\rm DM}\left(\frac{M_{\rm P}}{M_{\rm in}}\right)^{3/2},
\end{equation}which is the dark matter yield due to Hawking evaporation during a radiation-dominated epoch.

\paragraph{Matter-dominated era} The matter and radiation scale differently with temperature, note that for $\beta$ larger than a critical value $\beta_{\rm c}$, an early matter-dominated era is unavoidable. At an equality temperature $T_{\rm eq}$, we have $\rho_{\rm BH}(T_{\rm eq}) = \rho_{\rm rad}(T_{\rm eq})$, and hence \cite{Bernal:2021yyb},
\begin{equation}\label{eq:Teq}
    T_{\rm eq} = \beta\, T_{\rm in} \left( \frac{g_{\star, s}(T_{\rm in})}{g_{\star, s}(T_{\rm eq})} \right)^{1/3}.
\end{equation}

Therefore, the universe expansion will be driven by PBHs at temperatures lower than $T_{\rm eq}$. Naturally, it happens for $\rho_{\rm BH} > \rho_{\rm rad}$, that is, $\beta > \beta_{\rm c}$, where,
\begin{align}
    \beta_{\rm c} \equiv \frac{T_{\rm ev}}{T_{\rm in}}.
\end{align}

Thus, when the SM radiation reaches a critical temperature ($T_c$), the universe will not be a free fluid anymore, but rather driven by entropy injection due to PBH evaporation products with,  \cite{Bernal:2021yyb, Bernal:2021bbv} 
\begin{equation}
    T_{\rm c} \simeq \left(\frac{g_\star (T_{\rm in}) \pi}{5760} \frac{M_{\rm P}^{10} T_{\rm eq}}{M_{\rm in}^6}\right)^{1/5} \simeq (T_{\rm eq}\,T_{\rm ev}^4)^{1/5},
\end{equation}

During this time, the SM radiation redshifts with as $\rho_{\rm rad} \propto a^{-3/2}$, since $T(a) \propto a^{-3/8}$ instead of the usual $T(a) \propto a^{-1}$, where $a$ is the scale factor. 

Assuming instantaneous thermalization, the evaporation products will reach an equilibrium temperature $T_{\rm RH}$. We follow similar steps to determine the dark matter yield today , 
\begin{equation}
    Y^{\rm MD}_{\rm DM} = \frac{n_{\rm DM}(T_0)}{s(T_0)} = \frac{n_{\rm DM}(T_{\rm RH})}{s(T_{\rm RH})} =  N_{\rm DM} \frac{n_{\rm BH}(\tau)}{s(T_{\rm RH})},
\end{equation}
In a matter-dominated universe, 
we can equate the Hubble parameter to the Friedmann equation for $\rho_{\rm BH}$ driving the expansion rate to get,
\begin{equation}
   n_{\rm BH}(\tau) = \frac{\pi^2 g_{\star}(T_{\rm in})^2 }{19200}\frac{M_{\rm P}^{10}}{M_{\rm in}^7}.
\end{equation}
Assuming $s(T_{\rm RH}) = s(T_{\rm ev})$ and $g_{\star, s}(T)= g_{\star}(T)$ here, we can write the dark matter yield as, \cite{Gondolo:2020uqv}
\begin{equation}\label{eq:yieldmat}
    Y_{\rm DM}^{\rm MD} = \frac{3}{8}\left(\frac{g_{\star}(T_{\rm ev})}{40} \right)^{1/4} \left(\frac{M_{\rm P}}{M_{\rm in}}\right)^{5/2} N_{\rm DM},
\end{equation}

Note that neither the yield nor the relic abundance depend on $\beta$, because the PBH evaporation products determine the entropy density at the evaporation time.

The corresponding relic density in both cases (radiation-domination or matter-domination) can be obtained by substituting Eq.~\eqref{eq:yieldrad} and Eq.~\eqref{eq:yieldmat} into Eq.~\eqref{eq:relicdensity} using Eqs.~\eqref{eq:Ntot1} and \eqref{eq:Ntot2}. For analytical equations, we refer to \cite{Gondolo:2020uqv}\footnote{We point out that we use the ``reduced'' Planck mass $M_{\rm P} = 1/\sqrt{8 \pi G}$, whereas they adopted the Planck mass $M_{\rm Pl} = 1/\sqrt{G}$.}. 

We have derived the dark matter yield due to Hawking radiation in matter-domination and radiation-domination regimes.  We will now consider the presence of an additional DM population, represented by a fermion ,$\chi$, produced via freeze-out in a non-standard cosmology due to PBH. In the next section, we will review dark matter production in non-standard cosmologies and then connect it with the PBHs. 

\subsection{\label{sec:freezeout}Freeze-out mechanism}

Assuming the dark matter particles were in thermal equilibrium at early due to sizeable interactions with SM particles, we arrive at the conclusion that eventually they decoupled because
the expansion rate equaled the interaction rate, making the interactions rarer up to the point of freeze-out, leading to a leftover abundance of dark matter particles in the universe. The evolution of the dark matter number density $n_{\chi}$ is described by the Boltzmann equation, 
\begin{equation}\label{eq:BEQfo}
   \frac{dY_{\chi}(x)}{dx} =  - \frac{s(x) \langle \sigma v \rangle }{x\,H(x)}\left[Y^2_{\chi}(x) - Y^{\textrm{eq}\,2}_{\chi}(x)\right]
\end{equation}
where $Y_{\chi} \equiv n_\chi / s$ and $x \equiv m_{\chi}/T$ is a ``time'' variable that helps us to simplify the integration and physical interpretations and the equilibrium comoving number density
\begin{equation}
     Y^{\textrm{eq}}_{\chi}(x) = \frac{45}{4\pi^4}\frac{g_{\chi}}{g_{\star s}}x^2 K_2 (x),
\end{equation}
$g_{\chi}$ corresponds to the dark matter degrees of freedom and $K_2(x)$ is the modified Bessel function. Moreover, we rewrote the entropy density in terms of the new parameter,
\begin{equation}
    s(x) = \frac{2\pi^2}{45}g_{\star s}(x)m^3_{\chi} x^{-3}.
\end{equation}
In the standard lore, dark matter decouples during an epoch of radiation domination. However, since we assume that PBH energy density eventually overcomes the SM radiation density before disappearing, we have to account for the fact that freeze-out may happen during different eras. According to the temperature parametrization described in the previous sections, the standard and non-standard scenarios can be distinguished as follows:
\begin{itemize}
    \item {\it standard freeze-out}, $T_{\rm ev} > T_{\rm fo}$: Dark matter decouples during a radiation-dominated era but after PBH evaporation. That is, whether or not PBHs come to dominate the energy density of the universe, their evaporation shall not affect the dark matter abundance. We dub this case as {\it standard freeze-out}.

    \item {\it early radiation-dominated freeze-out}, $T_{\rm fo} > T_{\rm eq}$: Again dark matter decouples during a radiation-dominated epoch but before PBH domination. Hence, PBH injects entropy into the SM plasma via Hawking evaporation, diluting dark matter number density. We dub it {\it early radiation-dominated freeze-out} (ERD).

    \item {\it early matter-dominated freeze-out}, $T_{\rm eq} > T_{\rm fo} > T_{\rm c}$: Since dark matter decoupling occurs during the PBH domination regime, still before it evaporates, we name this scenario {\it early matter-dominated freeze-out} (EMD). Therefore, the temperature is still inversely proportional to the scale factor, $T \propto a^{-1}$, since the PBHs are not yet efficiently evaporating into SM particles, i.e. they drive the Hubble expansion but not the temperature of the SM radiation plasma.

    \item {\it critical matter-dominated freeze-out}, $T_{\rm c} > T_{\rm fo} > T_{\rm ev}$: PBHs dominate the energy budget of the universe and govern the temperature evolution the SM radiation plasma, $T \propto a^{-3/8}$. We call it {\it critical matter-dominated freeze-out} (CFO) since it starts after $T_{\rm c}$.
    
\end{itemize}
The entropy injection into the thermal bath can be evaluated assuming energy conservation  before and after PBH evaporation,
\begin{equation}
    \frac{s(T')}{s(T_{\rm ev})} = \left(\frac{T'}{T_{\rm ev}}\right)^3 = \left( 1 + D\right)^{3/4},
\end{equation}
where $T'$ is independent of $\beta$ and stands for the SM plasma temperature after PBH evaporation occurs, and $D$ is the dilution factor,
\begin{equation}\label{eq:D}
    D \equiv \beta \frac{T_{\rm in}}{T_{\rm ev}} > 1.
\end{equation}

In what follows, we obtain the approximate analytical solutions for each scenario, assuming s-wave annihilation cross-section.

\paragraph{\label{sec:std}Standard freeze-out} We solve Eq.~\eqref{eq:BEQfo} using the Hubble parameter given by,
\begin{align}
    H_{\rm R}(x) = \pi \sqrt{\frac{g_\star(x)}{90}}\frac{m_{\chi}^2}{M_{\rm P}\, x^{2}}.
\end{align} 
The approximate analytical solution is,
\begin{equation}
       Y_{\chi}^{\rm STD} = \frac{15}{\sqrt{5 \pi g_\star}} \frac{x_{\rm fo}}{m_\chi M_{\rm P} \langle \sigma v \rangle},
\end{equation}
with $g_\star= 106.75$ at the time of freeze-out and $x_{\rm fo} \equiv m_\chi / T_{\rm fo} $ defines the freeze-out temperature $T_{\rm fo}$,
\begin{equation}
    x_{\rm fo} = \ln \left[ \frac{3}{2} \sqrt{\frac{5}{\pi^5 g_{\star}(x_{\rm fo})}} \, g_\chi m_\chi M_{\rm P} \langle \sigma v \rangle x_{\rm fo}^{1/2} \right].
\end{equation}

Using Eq.~\eqref{eq:relicdensity}, we find the relic abundance,
\begin{align}
\Omega_{\chi}^{\rm STD} h^2 & = 2.82 \times 10^8 m_{\chi}  Y_{\chi}^{\rm STD}.
\end{align}
We remark that we used \texttt{micrOMEGAs5.0} to obtain the results for this scenario \cite{Belanger:2018ccd}.

\paragraph{\label{sec:erd} Early radiation-dominated freeze-out} On the other hand, for an ERD scenario, the comoving number density undergoes a dilution,
\begin{equation}
    \Omega_{\chi}^{\rm ERD} h^2 = 2.82 \times 10^8 m_{\chi}  \frac{Y_{\chi}^{\rm STD}}{D}.
\end{equation}

\paragraph{\label{sec:emd}Early matter-dominated freeze-out} In this scenario, the Hubble parameter is rewritten as
\begin{align} 
    H(x) &= H_{\rm R}(x_{\rm eq}) \sqrt{\frac{g_{\star,s}(x)}{g_{\star,s}(x_{\rm eq})}} \left(\frac{x_{\rm eq}}{x}\right)^{3/2} \\
    &= \Lambda(x) \frac{m_\chi^2}{M_{\rm P}} \left(\frac{\beta}{x_{\rm in}\, x^3}\right)^{1/2},
\end{align}
where we have used Eq.~\eqref{eq:Teq}, and
\begin{equation}
    \Lambda (x)= \frac{g_{\star,s}(x_{\rm in})^{1/6} \pi }{g_{\star,s}(x_{\rm eq})^{2/3}} \sqrt{\frac{g_\star(x_{\rm eq})g_{\star,s}(x)}{90}},
\end{equation}
$x_{\rm in, eq} = m_\chi/T_{\rm in, eq}$, and $g_{\star, s}(x) = g_{\star, s}(x_{\rm fo}) = {\rm cte}$, the analytical approximate solution for the dark matter yield in a PBH-dominated epoch is
\begin{equation}
    Y_\chi^{\rm EMD} = \frac{3}{2}\Lambda'(x_{\rm fo})\sqrt{\frac{\beta}{x_{\rm in}}} \frac{x_{\rm fo}^{3/2}}{ m_\chi M_{\rm P} \langle \sigma v \rangle},
\end{equation}
with 
\begin{equation}
     \Lambda'(x_{\rm fo})= \frac{1}{2 \pi^{3/2}} \sqrt{\frac{45}{g_{\star, s}(x_{\rm fo}) g_{\star}(x_{\rm eq})}} \left(\frac{g_{\star, s}(x_{\rm eq})^{2}}{g_{\star, s}(x_{\rm in})^{1/2}}\right)^{1/3}.
\end{equation}
The freeze-out temperature is given by
\begin{equation}
     x_{\rm fo} = \ln \left[  \left(\frac{x_{\rm in}}{\beta}\right)^{1/2} \frac{g_\chi \langle \sigma v \rangle m_\chi M_{\rm P}}{\Lambda(x_{\rm fo})} \right],
\end{equation}
Note the dependence on $\beta$. The analytical approximation relic abundance, with the entropy dilution, is
\begin{equation}
  \Omega_{\chi}^{\rm EMD} h^2 =  2.82 \times 10^8 \, \frac{m_\chi\,Y^{\rm EMD}_{\chi}}{D}.
\end{equation}
Now, let's consider freeze-out during the critical evaporation period.

\paragraph{\label{sec:cfo}Critical matter-dominated freeze-out}

The sudden evaporation of the PBHs leads to a non-free SM radiation fluid. In this case, $Y=n/s$ is not a comoving quantity, and Eq.~\eqref{eq:BEQfo} is not valid anymore. We can rewrite it for $N_\chi (a) \equiv n_\chi \, a^3$,
\begin{equation}\label{eq:modBEQ}
    \frac{d N_\chi (a)}{da} = - \frac{\langle \sigma v \rangle}{a^4\,H(a)}[N_\chi^2(a) - N_\chi^{{\rm eq}\,2}(a)],
\end{equation} where the Hubble parameter is given as a function of the scale factor,
\begin{equation}
    H(a) = \frac{\pi}{3}\sqrt{\frac{g_\star(T_{\rm ev})^3}{10 g_\star(T)^2}}\frac{T_{\rm ev}^2}{M_{\rm P}} \left(\frac{a_{\rm ev}}{a}\right)^{3/2},
\end{equation} where $a_{\rm ev} \equiv a(T_{\rm ev})$. Therefore, plugging it into the Boltzmann equation in Eq.~\eqref{eq:modBEQ}, we can solve it to get an approximate analytical solution
\begin{equation}
    N_\chi^{\rm CFO} = \frac{\pi}{2} \sqrt{ \frac{g_\star(T_{\rm ev})^3}{10 g_\star(T)^2}}\frac{T_{\rm ev}^2}{\langle \sigma v \rangle M_{\rm P}}(a_{\rm ev}\, a_{\rm fo})^{3/2}.
\end{equation}
To find a cosmological invariant yield, we have to determine $Y_{\chi}^{\rm CFO} \equiv N_\chi^{\rm CFO}/ (s\, a^3)$. To obtain the denominator we use
\begin{equation}
    s\, a^3 = \frac{2\pi^2}{45} g_\star(T_{\rm ev}) \, T_{\rm ev}^3 \,a_{\rm ev}^3,
\end{equation}to get,
\begin{equation}
    Y_\chi^{\rm CFO} = \frac{45 \pi}{4\, g_\star(x_{\rm fo})}\sqrt{ \frac{g_\star(x_{\rm ev})}{10}} \frac{x_{\rm fo}^4}{m_\chi  M_{\rm P} \langle \sigma v \rangle x_{\rm ev}^{3}},
\end{equation}where, 
\begin{equation}
    \frac{a_{\rm fo}}{a_{\rm ev}} = \left( \frac{T_{\rm ev}}{T_{\rm fo}} \right)^{8/3}.
\end{equation}
Note that the dark matter yield does not depend on $\beta$. In this case, the freeze-out temperature can be written as,
\begin{equation}
    x_{\rm fo} = \ln \left[ \frac{3}{2} \sqrt{\frac{5}{\pi^5 g_\star(x_{\rm ev})}} \frac{g_\chi m_\chi M_{\rm P} \langle \sigma v \rangle  x_{\rm fo}^{5/2}}{x_{\rm ev}^2} \right].
\end{equation}
Therefore, taking a conservative approximation for the dilution factor, if dark matter decouples this time, the relic density can be analytically approximated to, 
\begin{equation}
    \Omega_{\chi}^{\rm CFO} h^2 = 2.82 \times 10^8 \, \frac{m_\chi\,Y^{\rm CFO}_{\chi}}{D}.
\end{equation}

Therefore, it is clear that again a dilution occurs and one can modify the theoretical predictions for the relic density, but the impact of PBH on dark matter production should be confronted with a series of constraints from cosmological observations which will be addressed in the next section.

\subsection{\label{sec:pbhconstraints}PBHs constraints} 
\begin{figure}[!ht]
    \centering
    \includegraphics[width=1\linewidth]{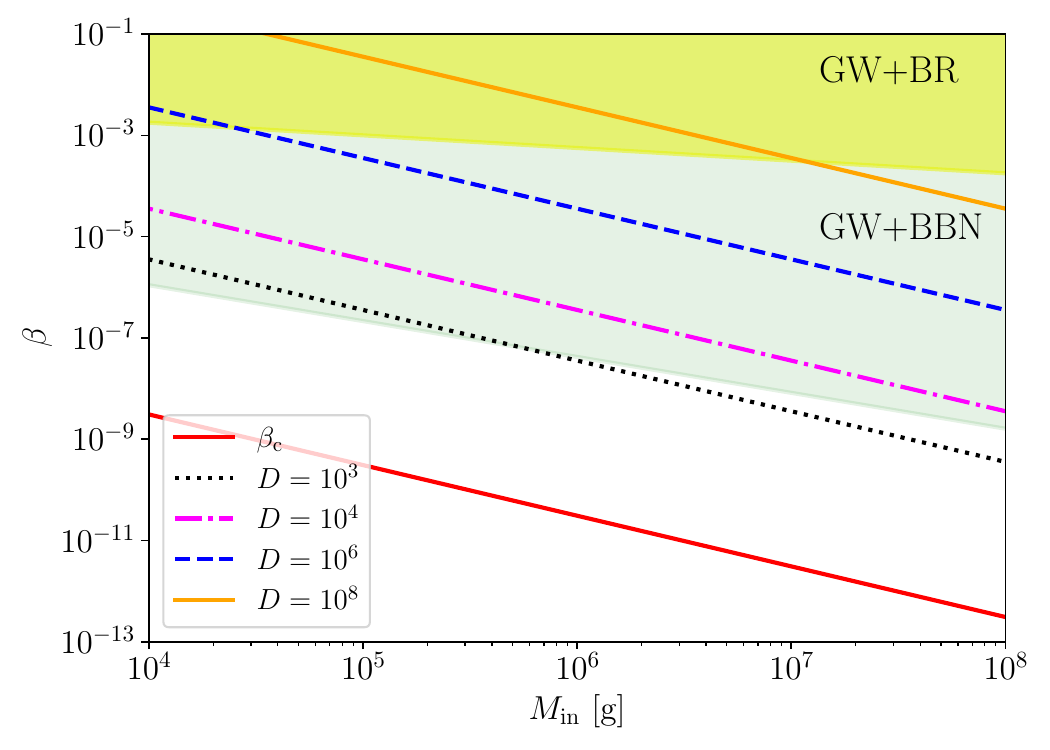}
    \includegraphics[width=1\linewidth]{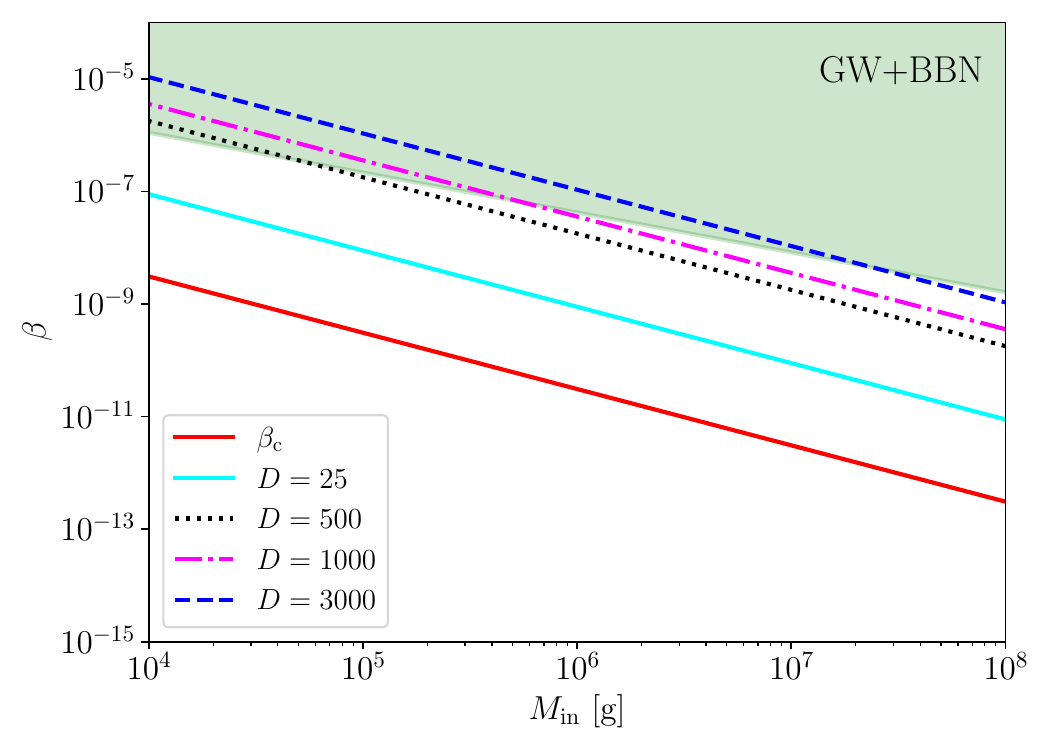}
    \caption{The upper panel shows the constraint of Eq.~\eqref{eq:GWbound} on the dilution factor was around $D \sim \mathcal{O}(10^{9})$ but when we take into account the enhancement due to energy stored within the scalar induced GWs produced during the instantaneous reheating phase, Eq.~\eqref{eq:BBNGWBounds}, it decreases to $D \sim \mathcal{O}(10^{3})$ orders of magnitude, as shown in the lower panel \cite{Inomata:2020lmk}.}
    \label{fig:betamin}
\end{figure}

The initial distribution $\beta$ and mass $M_{\rm in}$ are free parameters that we can use to determine the effects of PBHs in the early Universe. A portion of such a parameter space is constrained by cosmological observations. Cosmic Microwave Background (CMB) sets the Hubble rate during inflation $H_{\star} \leq 2.5 \times 10^{-5}\,M_{\rm P}$ \cite{Planck:2018vyg}\footnote{Note that they define $\sqrt{8 \pi} M_{\rm Pl} = 1$ (see their Fig. 15), hence, $M_{\rm Pl} = 1/\sqrt{8\pi G}$.}. From Eq.~\eqref{eq:Min}, we conclude that $M_{\rm in} \gtrsim 0.1$~g because lighter black hole can evaporate and significantly alter Planck's observations \cite{Gondolo:2020uqv}. Moreover, we enforce that PBHs must evaporate before the Big Bang Nucleosynthesis (BBN), $T_{\rm ev} > T_{\rm BBN} \simeq 4$~MeV, or $M_{\rm in} \lesssim 2\times 10^8$~g \cite{Arias:2019uol, Gondolo:2020uqv,Bernal:2020bjf}. Altogether, Summarizing, CMB and BBN tell us that
\begin{equation}\label{eq:CMBBBNBounds}
    0.1\,{\rm g} \lesssim M_{\rm in} \lesssim 2 \times 10^8\,{\rm g}.
\end{equation}
Besides the sensibility of CMB and BBN regarding radiation content, we also have to consider the possibility of the production of gravitational waves (GWs). Since we assume that PBHs eventually dominate the energy budget of the universe before evaporating, $\beta > \beta_{\rm c}$, the gravitational potential of these objects can induce GWs from second-order effects \cite{Papanikolaou:2020qtd}. If the induced GWs are too abundant, it can lead to a backreaction problem (BR), spoiling the current cosmological measurements. Therefore, to stay away from this scenario we constrain the initial abundance to be \cite{Papanikolaou:2020qtd}
\begin{equation}\label{eq:GWbound}
    \beta < 10^{-4}\left(\frac{10^9\,{\rm g}}{M_{\rm in}}\right)^{1/4},
\end{equation}
which ensures that the background energy will always be larger than the energy within the GWs. Furthermore, if we assume that during the transition from matter- to radiation-dominated era, the PBHs suddenly evaporate, there will be an enhancement in the induced GWs \cite{Inomata:2019ivs, Inomata:2020lmk}. Therefore, to make sure that this enhancement will not spoil BBN, the last bound becomes even stronger \cite{Domenech:2020ssp}
\begin{equation}\label{eq:BBNGWBounds}
    \beta < 1.1 \times 10^{-6} \left( \frac{\gamma}{0.2}\right)^{-1/2} \left(\frac{M_{\rm in}}{10^4\,{\rm g}}\right)^{-17/24}.
\end{equation}
For comparison, in Ref.~\cite{Bernal:2020bjf}, the authors set an upper bound on entropy dilution factor due to PBH evaporation to be $D \lesssim 10^{10}$, which is valid for Eq.~\eqref{eq:GWbound}. However, when we take the updated bound into account in Eq.~\eqref{eq:BBNGWBounds}, the dilution factor decreases to $D \sim \mathcal{O}(10^3)$, see Fig.~\ref{fig:betamin}. Therefore, it turns out to be the most stringent bound for an early matter-dominated scenario.

Other relevant constraints arise from capturing preexisting ultraheavy dark matter particles by PBHs before they evaporate \cite{Gondolo:2020uqv, Masina:2020xhk}. This possibility would occur if ultraheavy dark matter features an additional production mechanism such as gravitational production, for instance \cite{Chung:1998ua,Chung:1998zb,Chung:1998bt,Chung:2004nh}. However, as obtained in \cite{Gondolo:2020uqv}, and further discussed in \cite{Bernal:2020bjf}, it would imply $\beta \gtrsim 10^{-4}$ for $m_{\rm DM} \sim 10^{18}$~GeV, which is much weaker than Eq.~\eqref{eq:BBNGWBounds}. 

Besides capture, the requirement that the Hawking relics must be cold enough today not to disrupt small-scale structure observations leads to a limit over such a parameter space \cite{Fujita:2014hha, Auffinger:2020afu, Gondolo:2020uqv, Shallue:2024hqe}. However, as pointed out in Ref.~\cite{Bernal:2021akf}, it becomes relevant only for $m_{\rm DM} \lesssim 30$~GeV, prohibiting these masses mostly for $\beta > \beta_{\rm c}$. Moreover, as derived in Ref.~\cite{Shallue:2024hqe}, fermionic warm Hawking relics (keV-MeV) can only account for $\sim 2$\% of the total dark matter relics today. Nonetheless, these bounds do not apply for ultraheavy Hawking relics, $m_{\rm DM} \gtrsim 10^{13}\,{\rm GeV}\,({\rm g}/M_{\rm in})$, which are cold today. Therefore, for this study, we do not need to explicitly show these bounds. In the next section we will consider the thermal dark matter component introducing a specific Particle Physics framework to determine the conditions for freeze-out.

\section{\label{sec:models} The minimal $B-L$ model}

Despite our results being rather general, it is convenient to introduce a specific model to describe the thermal dark matter component dubbed $\chi$. We consider the minimal $B-L$ model to investigate the impact of the evaporating PBHs in the WIMP dark matter production. Besides the Standard Model (SM) gauge structure, we include the $U(1)_{B-L}$ Abelian symmetry, where $B$ and $L$ stand for the baryon and the lepton numbers, respectively. However, such a model is not anomally-free. Hence, we increase the particle content with three right-handed neutrinos $N_{iR}$ ($i=1,2,3$) to prevent gauge anomalies. Moreover, a complex scalar singlet $\Phi_S$ enters the game for breaking the $B-L$ symmetry giving rise to the $Z^\prime$ mass and leading to the Majorana mass term for the right-handed neutrinos that is a key element for generating active neutrino masses via type I Seesaw mechanism. Furthermore, we add a vector-like Dirac fermion $\chi$ to play the WIMP dark matter candidate, since after the $B-L$ spontaneous breaking, there remains a residual $Z_2$ symmetry under which only $\chi$ is odd. We summarize the particle content as well as their associated symmetry charges in Table ~\ref{tab:particlecontent}. At the end of the day, we only need three parameters to completely exploit the dark sector parameter space of the model: \{$g_{BL}, m_\chi, m_{Z^\prime}$\}.

\begin{table}
\begin{tabular}{| c | c | c | c | c | c |} 
 \hline 
\rule{0pt}{2.5ex}    
&  $SU(3)_{C}$ & $SU(2)_L$ & $U(1)_Y$ & $U(1)_{B-L}$ & $Z_{2}$ \\ \hline
\hline
\rule{0pt}{2.5ex}
 $q_{iL}$ & $\boldsymbol{3}$ & $\boldsymbol{2}$ & $1/6$ & $1/3$ & $+$\\ 
 \rule{0pt}{2.5ex}
 $u_{iR}$ & $\boldsymbol{3}$ & $\boldsymbol{1}$ & $2/3$ & $1/3$ & $+$ \\ \rule{0pt}{2.5ex}
 $d_{iR}$ & $\boldsymbol{3}$ & $\boldsymbol{1}$ & $-1/3$ & $1/3$ & $+$\\ \hline \rule{0pt}{2.5ex}
 $\ell_{iL}$ & $\boldsymbol{1}$ & $\boldsymbol{2}$ & $-1/2$ & $-1$ & $+$ \\ \rule{0pt}{2.5ex}
 $e_{iR}$ & $\boldsymbol{1}$ & $\boldsymbol{1}$ & $-1$ & $-1$ & $+$\\ \rule{0pt}{2.5ex}
 $N_{iR}$ & $\boldsymbol{1}$ & $\boldsymbol{1}$ & $0$ & $-1$& $+$ \\ \hline \rule{0pt}{2.5ex}
 $H$  & $\boldsymbol{1}$ & $\boldsymbol{2}$ & $-1/2$ & $0$& $+$ \\ \rule{0pt}{2.5ex}
 $\Phi_S$  & $\boldsymbol{1}$ & $\boldsymbol{1}$ & $0$ & $2$& $+$ \\ \hline 
 $\chi$ & $\boldsymbol{1}$ & $\boldsymbol{1}$ & $0$ & $1/3$& $-$ \\ \hline 
\end{tabular}
\caption{\label{tab:particlecontent} The particle content of the minimal B-L model according to their associated symmetry charges.}
\end{table}

The scalar Lagrangian that describes the $B-L$ symmetry breaking, generating the $Z^\prime$ mass, is then
\begin{equation}
    \mathcal{L}_{\textrm{S}} = (D_{\mu}\Phi_S)^{\dagger}(D^{\mu}\Phi_S) - \mu_{S}^2\Phi^{\dagger}_S \Phi_S - \frac{\lambda_S}{2} \left(\Phi^\dagger_S \Phi_S \right)^2, \label{eq:lscalar}
\end{equation}
where $\lambda_S$ is the scalar self-coupling, and we parametrize the field as follows
\begin{equation}
    \Phi_S = \frac{1}{\sqrt{2}}\left( v_S + \phi_S + i\rho_S \right),
\end{equation}
with $v_S$ standing for the $B-L$ vacuum expectation value (VEV), and $\rho_S$ is the respective Goldstone field. The Yukawa Lagrangian for describing the type I Seesaw mechanism is
\begin{equation}
    \mathcal{L}_{\textrm{Yukawa}} =  y_{ij}^{D}\overline{L}_i \Tilde{H}N_{jR} + y_{ij}^{M} \overline{(N^{C}_{iR})}\Phi_S N_{jR}, \label{eq:lyukawa}
\end{equation}
where $y_{ij}^{D, M}$ are the Dirac and Majorana Yukawa couplings, $\Tilde{H} = i\sigma_2 H$ is the isospin transformation of the SM Higgs doublet $H = \left(\phi^+, \phi^0\right)^{T}$, and $L_i$ is the lepton doublet for each $i$th-generation. The Seesaw mechanism is automatically implemented after the $B-L$ breaking, which takes place at high-energy scales $v_S \gg v$, with $v$ standing for the SM VEV. So, the tree-level mass of the scalar singlet, and the masses of the $Z^\prime$ and right-handed neutrinos are
\begin{align}
    m_{\phi_S} & = \sqrt{\lambda_S}\,v_S, \\
    m_{Z^\prime} &= 2g_{BL} v_S, \\
    m_{N_{iR}} &= \frac{y^{M}_{i}}{\sqrt{2}}v_S. \label{eq:BLmasses}
\end{align}
Notice that the Majorana mass matrix is diagonal, $i=j$. The details of the dark matter phenomenology are encoded in the Lagrangian 
\begin{align}
    \mathcal{L}_{\textrm{DM-SM}} &= i\overline{\chi}\gamma^{\mu}\partial_\mu\chi - m_\chi \overline{\chi}\chi \nonumber\\
    &- \frac{1}{4}F^{\prime\,\mu\nu}F^\prime_{\mu\nu} + g_{BL}n_\chi\overline{\chi}\gamma^{\mu}\chi Z^{\prime}_\mu \nonumber \\
    &+g_{BL} n_{f} \sum_{f} \overline{\psi}_f\gamma^{\mu}\psi_f Z^{\prime}_{\mu} \label{eq:ldm}
\end{align}
where $m_\chi$ is the dark matter mass, $g_{BL}$ is the $B-L$ gauge coupling, $n_\chi$ and $n_f$ are the dark matter and other fermions $B-L$ charges, respectively. By the way, $n_\chi$ must differ from $\pm 1$, otherwise, dark matter could decay due to a new Yukawa term concerning $\chi_R$. Furthermore, $F^{\prime}_{\mu\nu}$ is the strength tensor related to the $Z^\prime$ boson.

\begin{figure}
    \centering
    \includegraphics[width=1\linewidth]{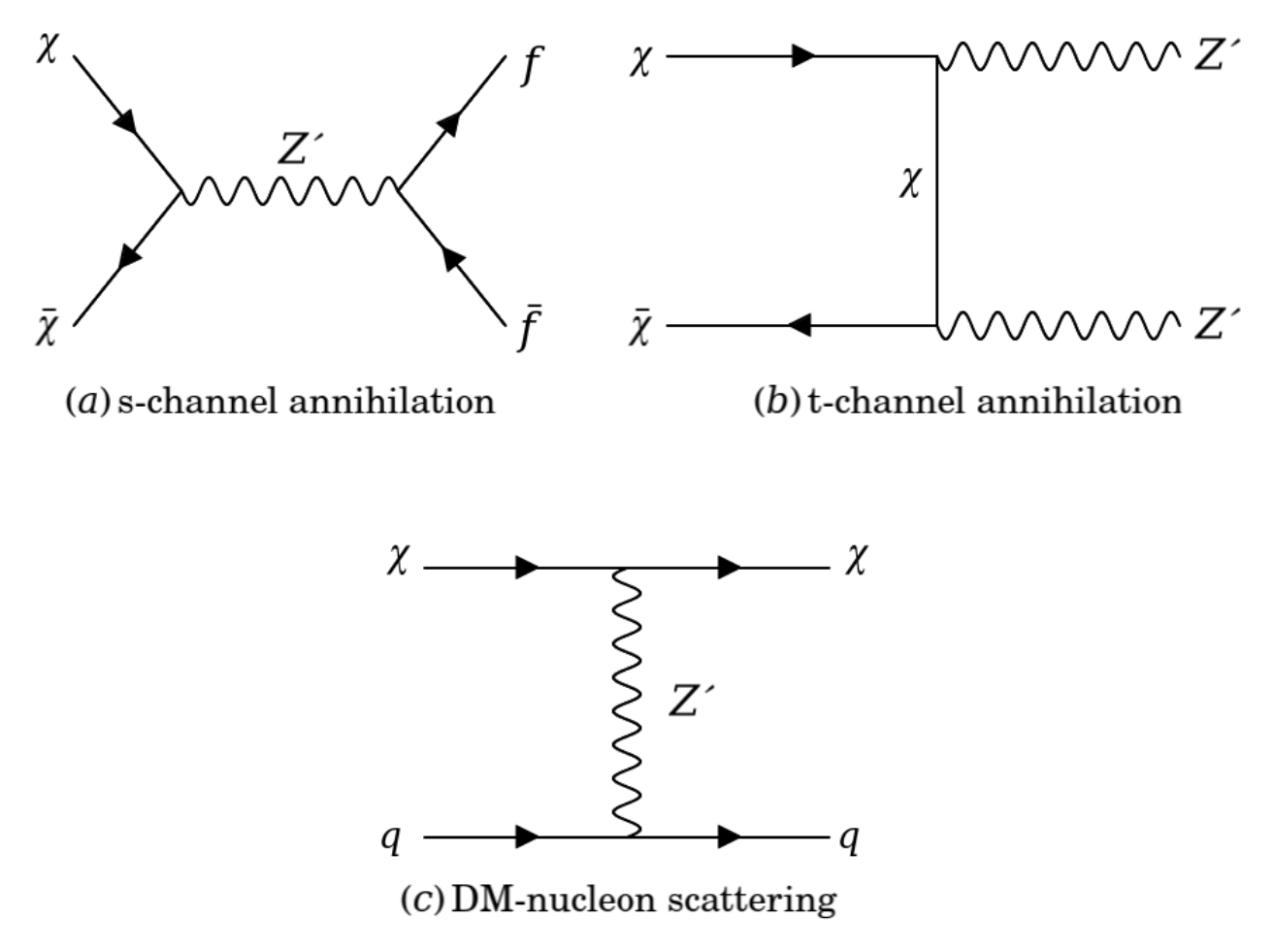}
    \caption{The Feynman diagrams for dark matter phenomenology: ($a$) dark matter annihilation into SM fermions, which is more relevant for $m_\chi < m_{Z^{\prime}}$ regime; ($b$) dark matter annihilation into on-shell $Z^\prime$, which is only open in $m_\chi > m_{Z^{\prime}}$ regime; and ($c$) dark matter-nucleon scattering process. The ($a$) and ($b$) diagrams apply for computing the relic abundance, while ($c$) applies for direct detection searches.}
    \label{fig:feynmandiagrams}
\end{figure}

\begin{figure*}
    \centering
    \includegraphics[width=0.45\linewidth]{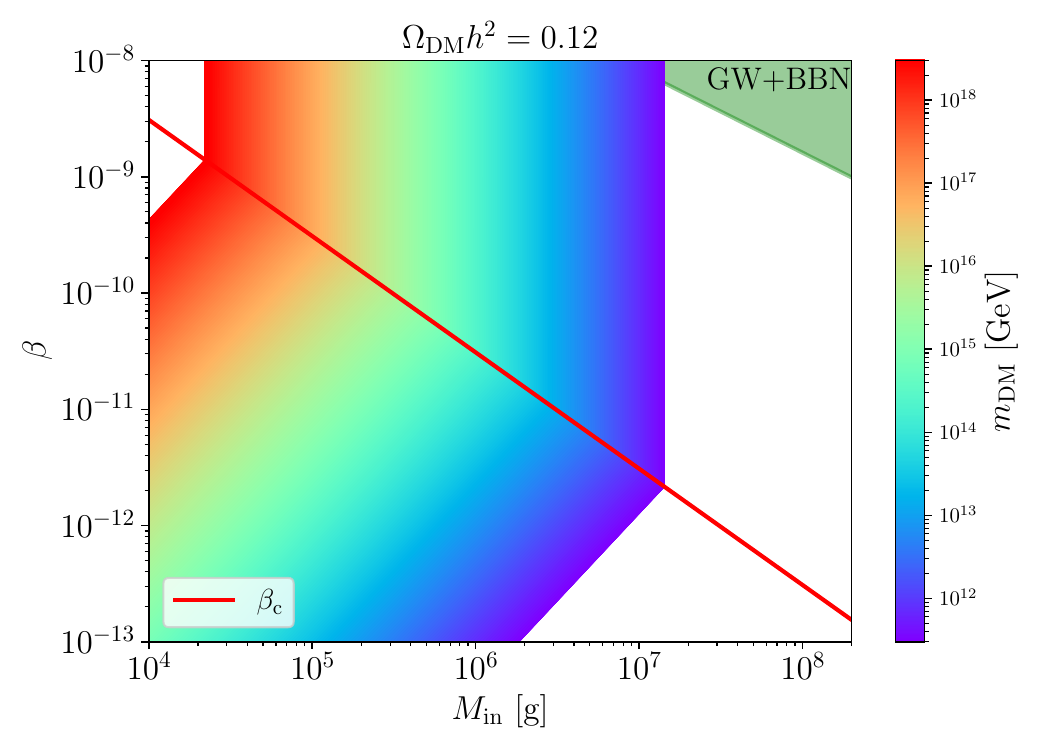}
    \includegraphics[width=0.45\linewidth]{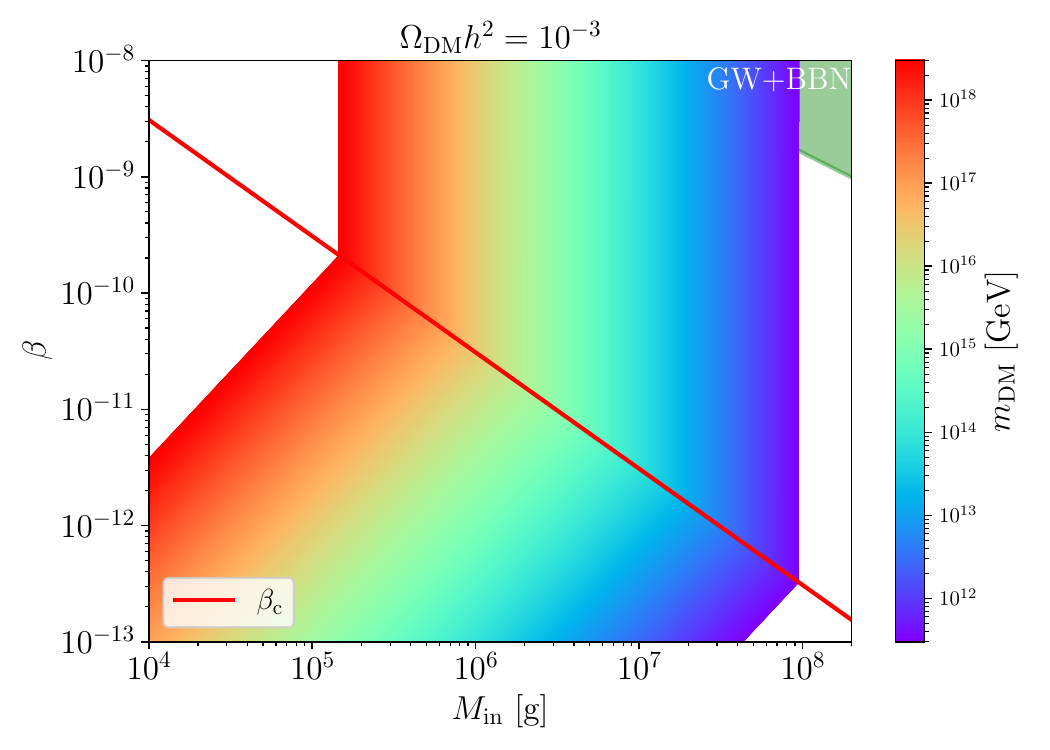}
    \caption{Dirac dark matter relic abundance produced by PBH evaporation. The green region indicates the GW bounds together with BBN. The red line divides the plot into two parts: for $\beta < \beta_{\rm c}$, it relies on the radiation-dominated part, and for $\beta > \beta_{\rm c}$ it gives the PBH-dominated one. The rainbow region, from violet to red, expresses the correct Hawking relic from $3\times10^{11}\,{\rm GeV} \leq m_{\rm DM} \leq 3\times10^{18}\,{\rm GeV}$. These panels demonstrate that ultraheavy dark matter particles are permitted, regardless of whether PBHs dominate or not, and whether they make up the entire dark matter relic today or not.}
    \label{fig:hawkingeva}
\end{figure*}

\begin{figure*}[!ht]
    \centering
    \begin{subfigure}[b]{0.45\linewidth}
        \centering
        \includegraphics[width=\linewidth]{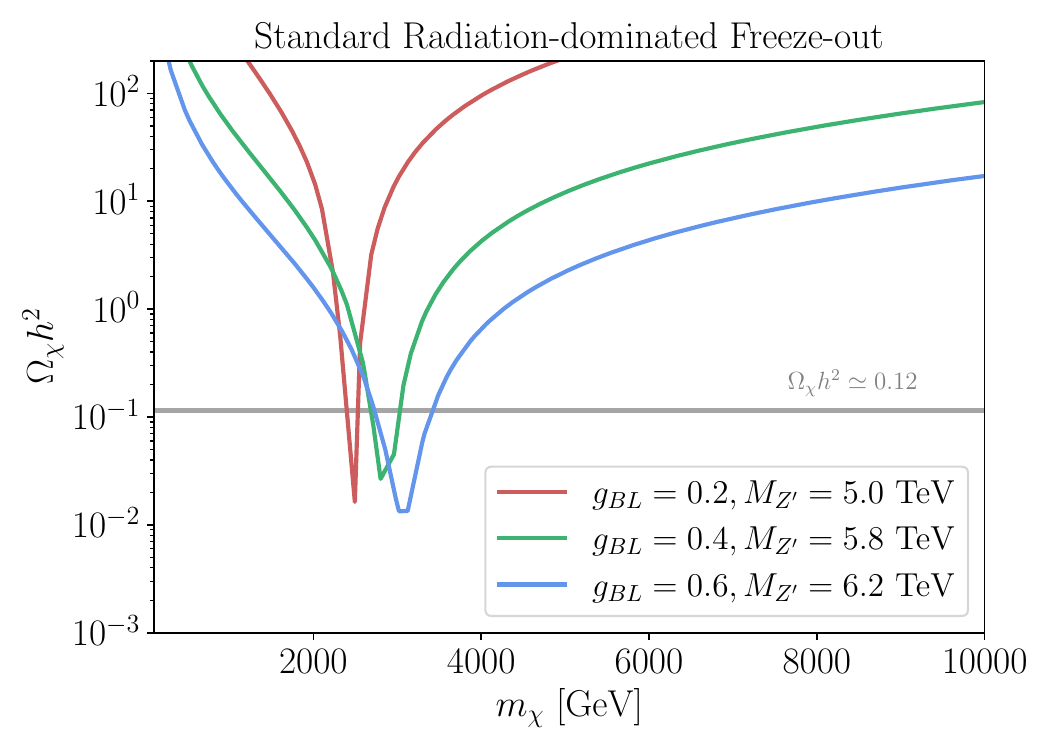}
        \caption{Standard radiation-dominated freeze-out}
        \label{fig:std}
    \end{subfigure}
    \hfill
    \begin{subfigure}[b]{0.45\linewidth}
        \centering
        \includegraphics[width=\linewidth]{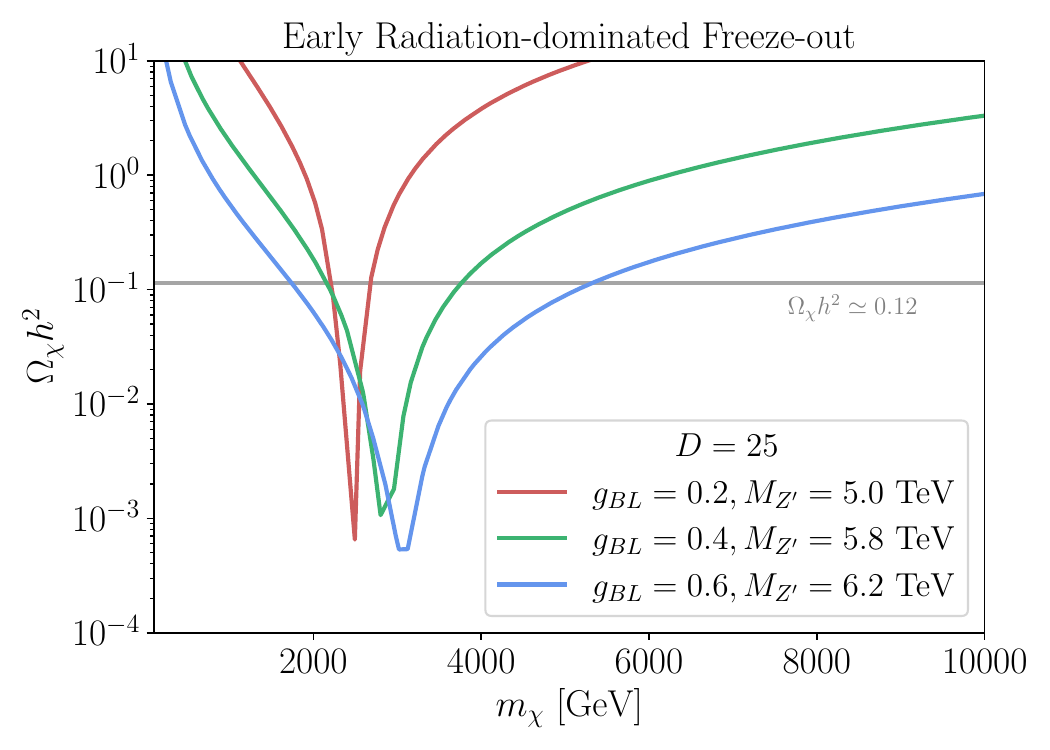}
        \caption{Effect of late entropy injection}
        \label{fig:erd}
    \end{subfigure}
    \vfill
    \begin{subfigure}[b]{0.45\linewidth}
        \centering
        \includegraphics[width=\linewidth]{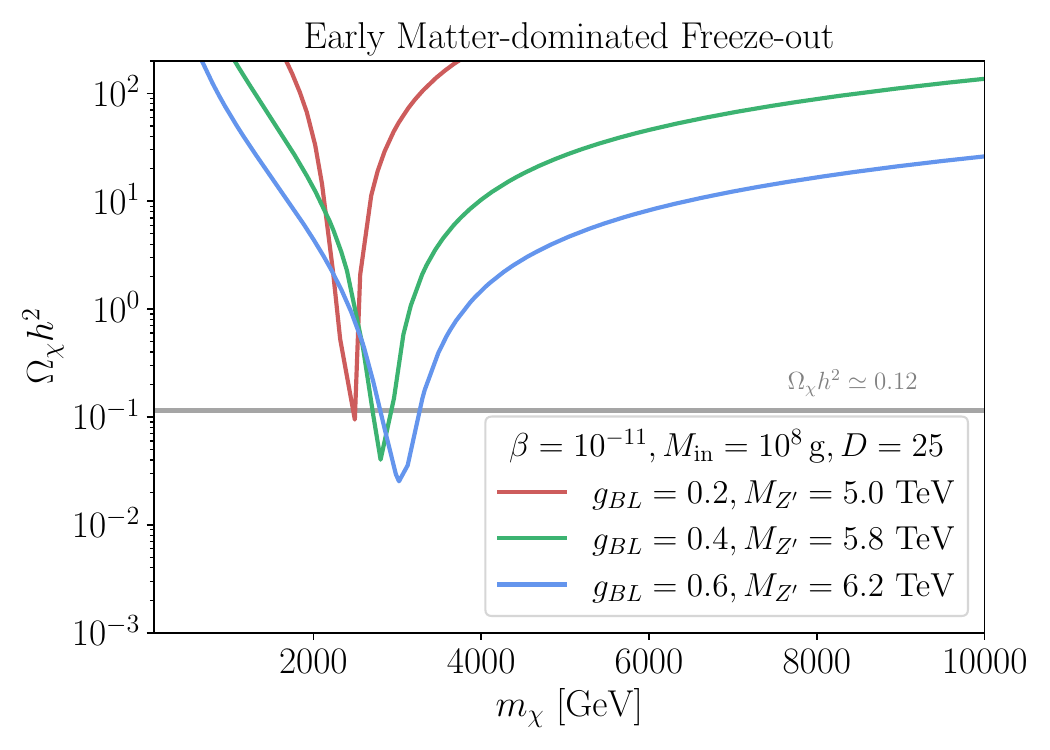}
        \caption{Freeze-out during PBH-dominated universe (free SM radiation fluid)}
        \label{fig:emd}
    \end{subfigure}
    \hfill
    \begin{subfigure}[b]{0.45\linewidth}
        \centering
        \includegraphics[width=\linewidth]{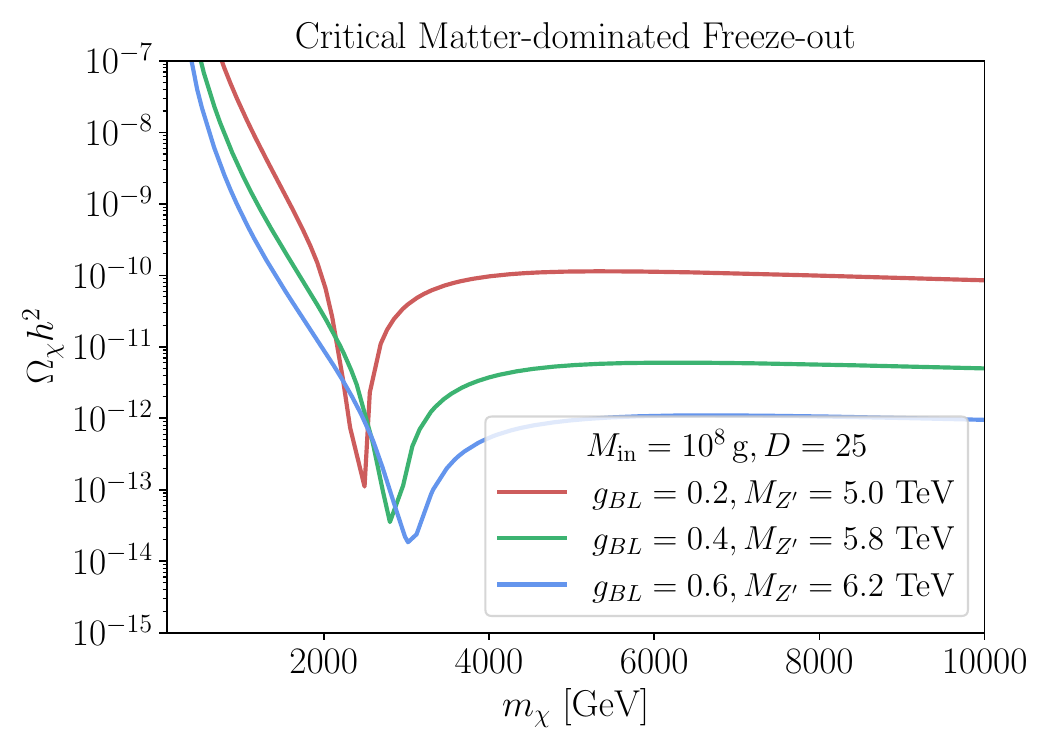}
        \caption{Freeze-out during PBH-dominated universe (sudden evaporation period)}
        \label{fig:cfo}
    \end{subfigure}
    \caption{We are presenting the dark matter relic abundance for various scenarios involving gauge couplings ($g_{BL}$) and free $Z'$ masses that are not constrained by collider bounds. The top panels display the results of freeze-out during the radiation-dominated era: (a) illustrates standard radiation-dominated freeze-out, and (b) demonstrates the effect of late entropy injection on the dark matter yield. The lower panels depict dark matter freeze-out during the PBH-dominated universe: (c) corresponds to freeze-out when the SM radiation plasma is a free fluid, and (d) corresponds to freeze-out when the thermal bath is driven by entropy injection during a sudden evaporation period.}
    \label{fig:omegaxmdm}
\end{figure*}

\begin{figure*}[!ht]
\begin{subfigure}[b]{0.45\linewidth}
       \centering
        \includegraphics[width=\linewidth]{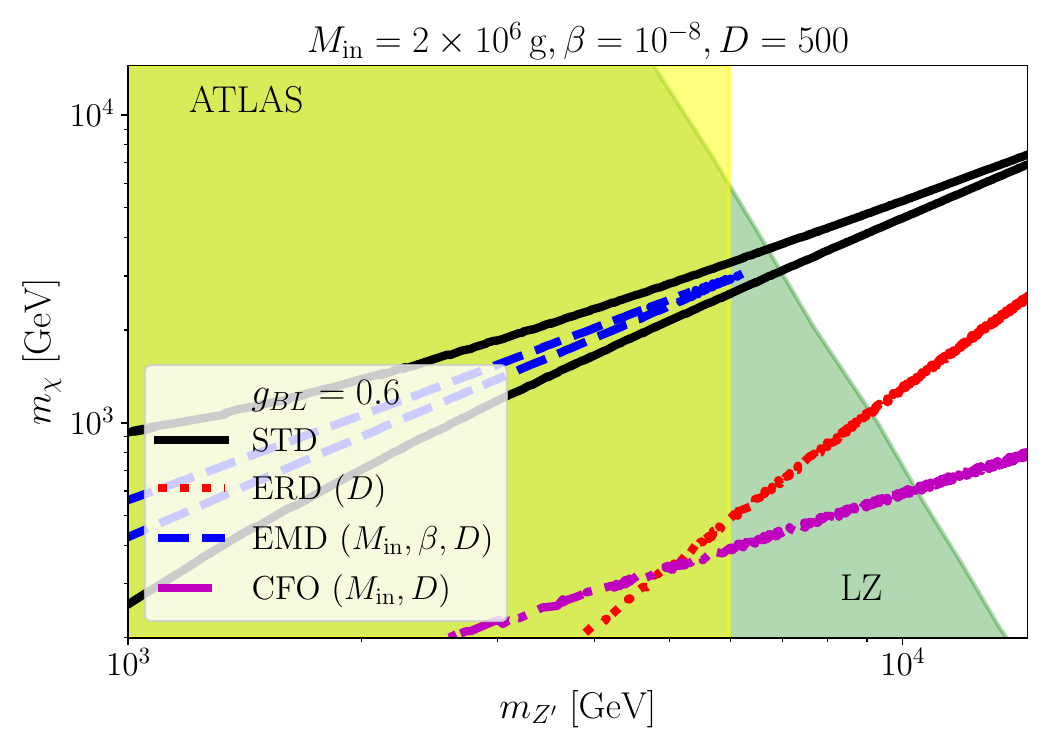}
        \caption{Thermal dark matter prospects for a PBH monochromatic initial mass $M_{\rm in} = 2 \times 10^6$~g, initial abundance $\beta = 10^{-8}$, and dilution factor $D = 500$.}
        \label{fig:Min1e6}
    \end{subfigure}
    \hfill
    \begin{subfigure}[b]{0.45\linewidth}
        \includegraphics[width=\linewidth]{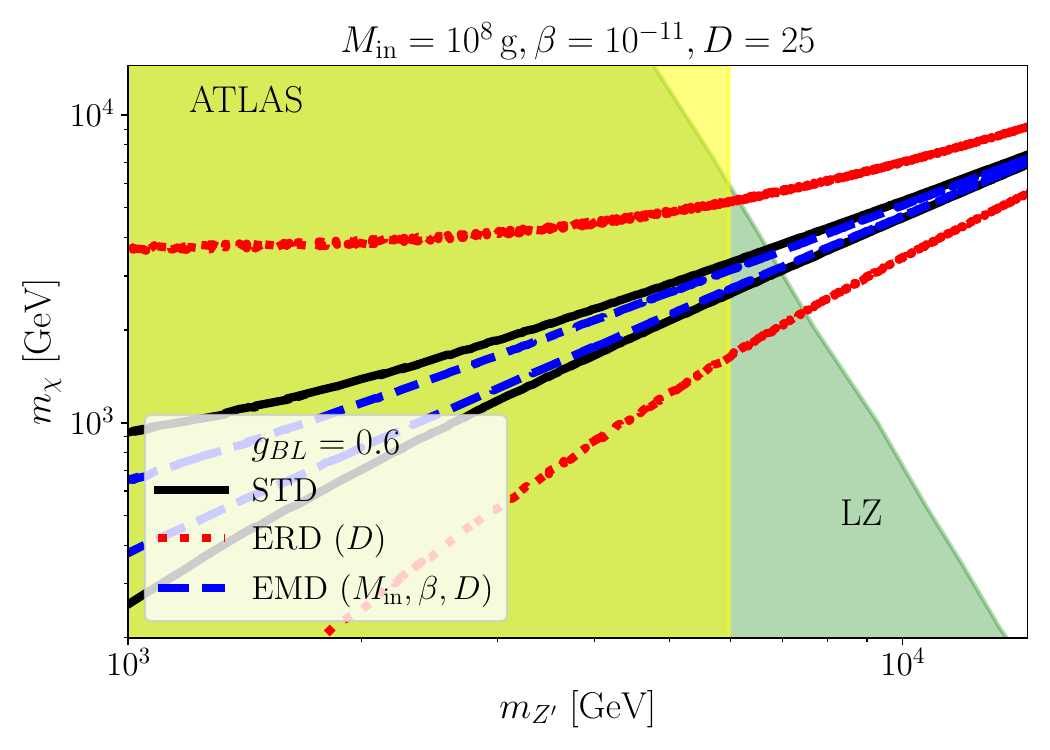}  
        \caption{Thermal dark matter prospects for a PBH monochromatic initial mass $M_{\rm in} = 10^8$~g, initial abundance $\beta = 10^{-11}$, and dilution factor $D = 25$.}
        \label{fig:Min1e8}
    \end{subfigure}
    \hfill
    \begin{subfigure}[b]{0.45\linewidth}
        \includegraphics[width=\linewidth]{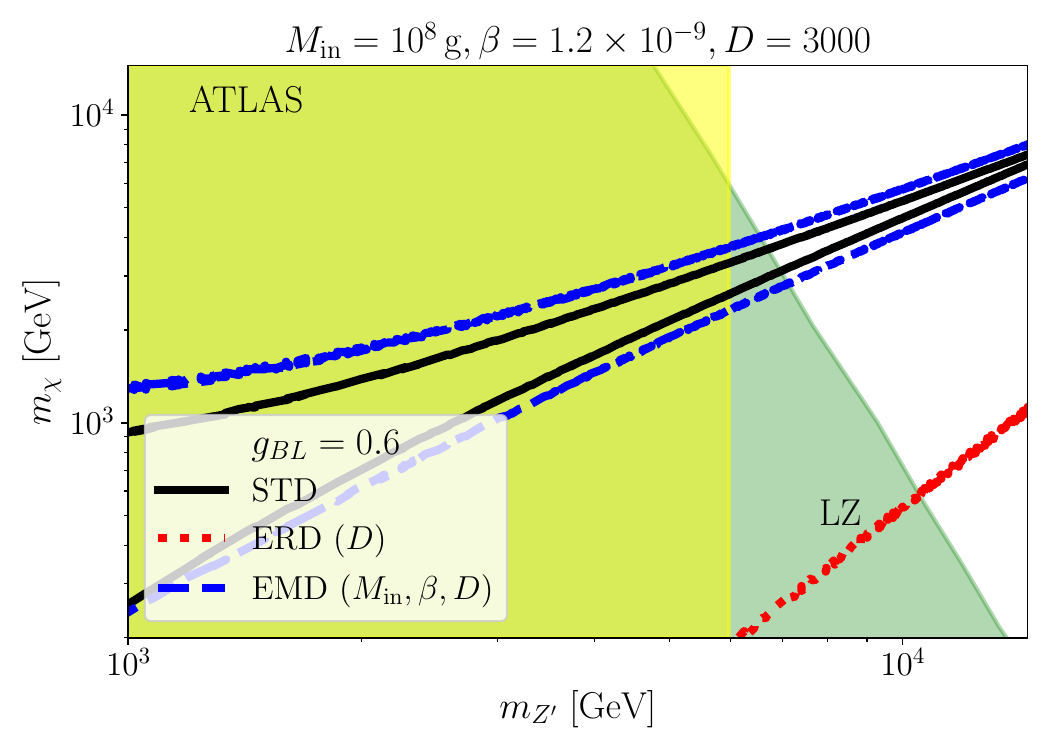}  
        \caption{Thermal dark matter prospects for a PBH monochromatic initial mass $M_{\rm in} = 10^8$~g, initial abundance $\beta = 1.2 \times 10^{-9}$, and the around highest allowed dilution factor $D = 3000$.}
        \label{fig:Min1e8higherD}
    \end{subfigure}
    \caption{The correct dark matter relic abundance for all freeze-out scenarios. The black contour is the result of the standard scenario. The red dotted line gives the ERD scenario. The dashed blue and dashed-dotted magenta lines are the freeze-out scenario during PBH domination, the former is the EMD case and the latter is the CFO case.}
    \label{fig:omega_g6}
\end{figure*}

We show the relevant Feynman diagrams for dark matter phenomenology in  Fig.~\ref{fig:feynmandiagrams}. Diagrams ($a$) and ($b$) enter the thermally-averaged annihilation cross-section $\langle \sigma v_{\textrm{rel}} \rangle_\textrm{ann}$ in the Boltzmann equation for computing the dark matter relic density. When the dark matter mass is smaller than the mediator mass, $m_\chi < m_{Z'}$, the s-channel is more relevant than the t-channel, and the annihilation cross-section, up to first-order on the relative velocity $v_{\textrm{rel}}$, is 
\begin{align}
    \langle \sigma v_{\textrm{rel}} \rangle_{\chi\overline{\chi} \rightarrow f \overline{f}} = &\frac{g^4_{BL}n^2_{\chi}n^2_{f}}{2\pi} \nonumber\\
    & \times \sum_{f}n_{c}^\textrm{f} \frac{\left( m_f^2 + 2m_{\chi}^2\right) \sqrt{1 - \frac{m^2_f}{m_\chi^2}}}{\left( m_{Z'}^2 - 4m_\chi^2\right)^2 + \Gamma^2_{Z'}m_{Z'}^2},
\end{align}
where $m_f$ and $n^\textrm{f}_c$ are the mass and the number of colors of the final state SM fermion, respectively. Moreover, $\Gamma_{Z^\prime}$ is the $Z^\prime$ decay width given by
\begin{align}
    \Gamma_{Z^\prime} &= \frac{g_{BL}^2 m_{Z^\prime} }{12 \pi S} \Bigg[ \sum_{f} n^{\textrm{f}}_c \left(1+2\frac{m^2_{f}}{m^2_{Z^\prime}}\right) \sqrt{1 - \frac{4 m_{f}^2}{m^2_{Z^\prime}}} \nonumber\\
    &+  \left(1+2\frac{m^2_{\chi}}{m^2_{Z^\prime}}\right) \sqrt{1 - \frac{4 m_{\chi}^2}{m^2_{Z^\prime}}} \,\Bigg],
\end{align}
where $S$ is a symmetry factor. However, once the dark matter mass becomes greater than the mediator mass, $m_\chi > m_{Z'}$, the ($b$)-diagram also contributes to the dark matter relic density. Again, considering only up to first-order expansion terms in $v_{\textrm{rel}}$, we have
\begin{align}
        \langle \sigma v_{\textrm{rel}} \rangle_{\chi\overline{\chi} \rightarrow Z' Z'} = \frac{g^4_{BL}n^4_{\chi}\left(m_{\chi}^2 - m_{Z'}^2  \right)^{3/2}}{4\pi m_{\chi}  \left( m_{Z'}^2 - 2m^2_{\chi}\right)^2}.
\end{align}
Finally, the third diagram is relevant for direct detection searches. For the $B-L$ model, only the spin-independent scattering cross-section matters since the $Z^{\prime}$ boson features vector coupling with both quark and dark matter pairs, which is
\begin{equation}
   \sigma_{\chi N=p,n}^{\rm SI} = \frac{\mu_{\chi N}^2}{\pi}\frac{9 \, n_q^2 n_\chi^2 g_{BL}^4}{m_{Z^{\prime}}^4},\label{eq:csSI}
\end{equation}
where $\mu_{\chi N}=\frac{m_\chi m_N}{m_\chi+m_N}$ is the WIMP-nucleon reduced mass, $N$ is the atomic number, and $p$ and $n$ stand for the proton and the neutron, respectively.

In a recent work \cite{Arcadi:2023lwc}, we have derived the current constraints to this model. On the collider side, we adopted the ATLAS dileptons limit \cite{ATLAS:2019erb}, since it is stronger than the ATLAS dijets \cite{ATLAS:2019fgd} and di-top \cite{ATLAS:2019npw}. It is important to note that the ATLAS collaboration has not looked for a $B-L$ $Z^\prime$ boson. Therefore, we had to establish the limits for the model ourselves. We compared our theoretical results for the $pp \to Z' \to f \bar f$ process with the fiducial cross-section times branching ratio obtained by ATLAS at a 95\% confidence level \cite{ATLAS:2019erb}. For each pair of $B-L$ gauge coupling and $Z'$ mass, we used {\tt CalcHEP} to calculate the $Z'$ branching ratio into leptons, and together with the {\tt MadGraph5} we performed the Monte Carlo simulation using the NNPDF23LO parton distribution function to calculate the complete event $\sigma(pp \to Z^{\prime}) \times \Gamma (Z' \to \ell\bar{\ell})$ at  $\sqrt{s}=13$ TeV, with $\ell = e,\mu$ \cite{Belyaev:2012qa,Alwall:2014hca,Frederix:2018nkq,Carrazza:2013axa}. Besides requiring two opposite charge leptons for the signal events, we also took into account the same configuration of kinematic variables as the collaboration, say, transverse momentum $p_{T} > 30$~GeV and pseudorapidity $|\eta| <2.5$. Finally, we faced it against the ATLAS dilepton data \cite{ATLAS:2019npw}. In this work, we did not need to show the LEP-II bounds, since the limits for the parameter space we show in the \ref{sec:Results}~Results section still lie upon the center-of-mass energy of ATLAS \cite{Arcadi:2023lwc}. Moreover, xenon-based direct detection experiments, such as LUX-ZEPLIN (LZ) \cite{LZ:2022ufs} and XENONnT \cite{XENON:2023sxq}, strongly constrain the interaction of Eq.~\eqref{eq:csSI}. It is sufficient to show only the LZ limits in the dark matter mass range we are laying over. In Sec.~\ref{sec:Results}, we display the direct detection bounds obtained by comparing our WIMP-nucleon scattering cross-section in Eq.~\eqref{eq:csSI} against the LZ limits.

\section{\label{sec:Results} Results \& Discussion}

In Fig.~\ref{fig:hawkingeva}, we display the relic abundance of a Dirac dark matter resulting solely from PBH evaporation. We can see that Hawking evaporation can produce enough dark matter to explain Planck's data. The red line at $\beta = \beta_{\rm c}$ represents the transition from a radiation-dominated to a matter-dominated epoch. The green ones represent constraints from GWs, accounting for backreaction and ensuring a safe outcome for BBN after PBH's sudden evaporation regime. The chosen region of the parameter space agrees with CMB and BBN constraints. We focus on the $\beta > \beta_{\rm c}$ regime. The panels demonstrate that ultraheavy Hawking relics, $3 \times 10^{11}\,{\rm GeV} \leq m_{\rm DM} \leq 3\times10^{18}\,{\rm GeV}$, are permitted, regardless of whether they make up the entire dark matter relic or not. Since we want to exploit the effects of PBH domination in the parameter space of the WIMP candidate, we take the ultraheavy Hawking relics to account only for a tiny fraction of the total observed abundance. 

Fig.~\ref{fig:omegaxmdm} shows the relic density as a function of the WIMP dark matter mass $m_\chi$ for the four different regimes described in Sec.~\ref{sec:freezeout}. The values of the gauge couplings and $Z'$ masses were chosen accordingly to collider bounds. The (a) panel shows the standard scenario, while (b) shows the effect of late entropy injection on the thermal yield. The (c) panel represents dark matter freeze-out during the PBH-dominated universe with SM radiation plasma scaling as a free fluid, $a \propto T^{-1}$. On the other lower side, (d) gives the critical freeze-out scenario that considers WIMP production during the sudden evaporation period, where $a \propto T^{-8/3}$. 

We choose the initial PBH abundance to be $\beta = 10^{-11}$, the initial mass $M_{\rm in} = 10^8$~g, and the dilution factor $D = 25$ accordingly to Eq.~\eqref{eq:BBNGWBounds}. Although small, the entropy injection alters the thermal relic. Especially in the ERD scenario, we can see the sensitivity of the widths to $D$ in Figs.~\ref{fig:std} and \ref{fig:erd}. It changes the order of magnitude and allows the viability for $m_\chi$ with the correct relic for wider values. Similarly, the dilution factor affects the EMD scenario in Fig.~\ref{fig:emd}. However, the orders of magnitude essentially do not change compared to the standard case because this case has a stronger dependence on $\beta$ and $M_{\rm in}$, whose values live close to $\beta_{\rm c}$. The most dramatic change is found in the CFO scenario in Fig.~\ref{fig:cfo}, where the entire parameter space is underabundant.

Concerning the WIMP dark matter phenomenology, Fig.~\ref{fig:omega_g6} summarizes our main results. The dark matter particle parameters are set as follows: We fixed $g_{BL} = 0.6$, scanning over the ($m_{Z'}, m_\chi$) parameter space. The yellow band is the ATLAS bound. The vertical line on a specific $Z'$ mass stands for the crossing point at the parameter space of $\sigma(s)$ {\it vs} $\sqrt{s}=m_{Z'}$, for a fixed $g_{BL}$, between our theoretical predictions and the ATLAS dilepton measurements. Hence, the ATLAS dilepton measurements exclude our model for $m_{Z'} \leq 5.97$~TeV. The green region corresponds to the limit from the LZ experiment. We can understand the shape of this limit using Eq.~\eqref{eq:csSI}, where it will not rely on $m_\chi$ for dark matter particles much heavier than the nucleus mass. Therefore, it goes as $\sigma^{\rm SI}_{\chi N} \propto (g_{BL}/Z')^4$. However, since we fixed the gauge coupling, it becomes $\sigma^{\rm SI}_{\chi N} \propto 1/Z'^4$.

We present the correct relic abundance for all freeze-out scenarios. For the cosmological parameters, we assume three different benchmark points: (a) $M_{\rm in} = 2 \times 10^6$~g, $\beta = 10^{-8}$, and $D=500$, which is displayed in Fig.~\ref{fig:Min1e6}; (b) $M_{\rm in} = 10^8$~g, $\beta = 10^{-11}$, and $D=25$, which is shown in Fig.~\ref{fig:Min1e8}; and (c) $M_{\rm in} = 10^8$~g, $\beta = 1.2 \times 10^{-9}$, and $D=3000$, which is presented in Fig.~\ref{fig:Min1e8higherD}. On one hand, the black curves give the observed abundance for the standard freeze-out scenario, i.e. when dark matter density does not undergo any effect due to PBHs. On the other hand, we have the non-standard scenarios: The dotted red curve results from the ERD case, the dashed blue curve stands for the EMD scenario, and the magenta gives the CFO freeze-out. The parentheses indicate the cosmological parameter dependence of each scenario. 

In Fig.~\ref{fig:Min1e6}, the EMD case is completely constrained, while the ERD and CFO scenarios provide new prospects for thermal dark matter masses in the future. In Fig.~\ref{fig:Min1e8}, the standard and EMD scenarios are indistinguishable, while the ERD case still allows for lighter masses compared to the standard scenario. Moreover, the CFO case does not appear because, on the particle physics side, it requires smaller annihilation cross-sections, i.e. heavier $Z'$ or smaller gauge couplings, and on the cosmological side, it demands lighter PBH masses. Fig.~\ref{fig:Min1e8higherD} shows the outcomes according to the highest permitted dilution factor. Interestingly, it presents the EMD regime going away from the standard case. It remarks on the complexity of playing with the three parameters. Nonetheless, the ERD regime keeps accessing smaller WIMP masses as larger as the dilution factor becomes, and again the CFO scenario can not survive to $M_{\rm in} = 10^8$~g.

The results indicate an open window for an interplay between Hawking evaporation and freeze-out production. Therefore, if PBHs led to a matter-dominated phase at early times, they gravitationally produced a tiny fraction of ultraheavy Hawking relics, and, parallelly, a WIMP candidate within the scope of a minimal $U(1)_{B-L}$ model freezes out before the complete evaporation of the PBHs. Especially, a monochromatic PBH distribution with initial mass $M_{\rm in} = 2 \times 10^6$~g supports freeze-out during ERD and CFO scenarios, while $M_{\rm in} = 10^8$~g can accommodate ERD and EMD freeze-out scenarios. It is worth highlighting that the standard freeze-out case is valid only for $\beta < \beta_{\rm c}$. However, since it does not undergo any effect from PBHs, we did not focus on this case.

\section{\label{sec:Conclusions} Conclusions}

Throughout this work, we have assumed that ultralight Schwarzschild's PBHs that arise from a monochromatic mass distribution eventually dominate the universe's energy density at early times, and evaporate before BBN. Hawking evaporation produces all accessible particle species, including SM radiation and ultraheavy dark matter. On top of that, a WIMP dark matter embedded in a minimal $B-L$ model is thermally produced via freeze-out during non-standard cosmological scenarios led by PBHs. These two mechanisms are for a two-component dark matter model, with the WIMP candidate as the most abundant component, while the ultraheavy Hawking relics contribute to a tiny fraction.

Due to the PBH-dominated period and consequent evaporation, an entropy injection into the SM radiation plasma is unavoidable, leading to a dilution of the thermal dark matter yield for all non-standard freeze-out scenarios. We have shown explicitly that such a dilution has to be at most of the order of $\mathcal{O}(10^3)$ to not spoil BBN due to the energy density stored in the induced GW produced by the PHBs. In this regard, an interplay between ultraheavy Hawking relics and early-radiation domination freeze-out is favorable for PBH initial masses around $2 \times 10^6$~g and $10^8$~g. The mixed scenario is also addressed for an early-matter domination case with $M_{\rm in} = 10^8$~g, although it behaves similarly to the standard case, in which PBHs never dominate. 

In other words, the relic density of thermal relics can be greatly affected by the existence of ultralight PHBs. Depending on the parameters of PHB physics new regions of parameter space open up, strengthening the continuous search for thermal relics signals in direct detection experiments and colliders. 

\paragraph*{Acknowledgements\,:} 
The authors thank Sven Fabian and David Cabo-Almeida for relevant and insightful discussions, and Óscar Zapata for an amazing trilogy of lectures at the International Institute of Physics (IIP) on Primordial Black Holes and Dark Matter. This work was supported by the Propesq-UFRN Grant 758/2023. The authors acknowledge the use of the IIP cluster ``{\it bulletcluster}". JPN acknowledges support from the Programa Institucional de Internacionalização (PrInt) and the Coordenação de Aperfeiçoamento de Pessoal de Nível Superior (CAPES) under the CAPES-PrInt Grant No. 88887.912033/2023-00. FSQ is supported by Simons Foundation (Award Number:1023171-RC), FAPESP Grant 2018/25225-9, 2021/01089-1, 2023/01197-4, ICTP-SAIFR FAPESP Grants 2021/14335-0, CNPq Grants 307130/2021-5, and ANID-Millennium Science Initiative Program ICN2019\_044.

\bibliographystyle{JHEPfixed.bst}
\bibliography{References.bib}

\end{document}